\documentclass[a4paper,12pt]{article}
\usepackage[dvipdfm]{graphicx}
\usepackage{graphicx}
\usepackage{amsmath}

\renewcommand{\dj}{d\kern-0.4em\char"16\kern-0.1em}
\renewcommand{\DJ}{\raise0.3ex\hbox{-}\kern-0.36em D}

\begin{document}

\title{EPR and magnetization studies on single crystals of a heterometallic (Cu$^{\mathrm{II}}$ and Cr$^{\mathrm{III}}$) complex: zero-field splitting determination} 
\author{Nikolina Novosel$^{\mathrm{a}}$, Dijana \v{Z}ili\'{c}$^{\mathrm{b,*}}$,  Damir Paji\'{c}$^{\mathrm{a}}$, Marijana Juri\'{c}$^{\mathrm{c}}$,\\ Berislav Peri\'{c}$^{\mathrm{c}}$, Kre\v{s}o Zadro$^{\mathrm{a}}$, Boris Rakvin$^{\mathrm{b}}$, Pavica Planini\'{c}$^{\mathrm{c}}$}
\date{}
\maketitle

\begin{center}
\textit{$^{\mathrm{a}}$Department of Physics, Faculty of Science, University of Zagreb, \\Bijeni\v{c}ka cesta 32, 10000 Zagreb, Croatia\\
$^{\mathrm{b}}$Division of Physical Chemistry, Ru\dj er Bo\v{s}kovi\'{c} Institute, \\Bijeni\v{c}ka cesta 54, 10000 Zagreb, Croatia\\
$^{\mathrm{c}}$Division of Materials Chemistry, Ru\dj er Bo\v{s}kovi\'{c} Institute, \\Bijeni\v{c}ka cesta 54, 10000 Zagreb, Croatia}
\end{center}

\begin{abstract}

Magnetic properties of single crystals of the heterometallic complex [Cu(bpy)$_3$]$_2$[Cr(C$_2$O$_4$)$_3$]NO$_3\cdot$9H$_2$O (bpy = 2,2'-bipyridine)
have been investigated. From the recorded EPR spectra, the spin-Hamiltonian parameters have been determined. The magnetization measurements have shown  magnetic anisotropy at low temperatures, which has been analysed as a result of the zero-field splitting of the Cr$^{\mathrm{III}}$ ion. By fitting the exactly derived magnetization expression to the measured magnetization data, the axial zero-field splitting parameter, $D$, has been calculated. Comparing to the EPR measurements, it has been confirmed that $D$ can be determined from the measurements of the macroscopic magnetization on the single crystals.

\end{abstract}

\section{Introduction}

In the past several years a great number of homo- and heteropolynuclear complexes containing different transition metal ions have been synthesized
and their magneto-structural properties have been investigated. Interaction between paramagnetic centres involved as well as the crystal structures of these compounds determine their magnetic properties. In describing magnetic behaviour of these species, besides the exchange interaction between metal ions, it is often necessary to include the ligand field effects which induce the anisotropy of paramagnetic centres and zero-field splitting. Paramagnetic metal centres of these complexes are usually bridged by diamagnetic ligands which mediate exchange interaction between metal ions \cite{kahn}. In that sense, a widely used ligand is the oxalate C$_2$O$_4^{2-}$ anion -- not only because of the influence it has in transmitting electronic effects between magnetic centres, but also because of its extraordinary rich binding facilities \cite{dva}. The use of stable mononuclear anionic oxalate complexes, [M$^{\mathrm{III}}$(C$_2$O$_4$)$_3$]$^{3-}$ (M$^{\mathrm{III}}=$ Cr, Fe, Ru), as ligands toward other metal ions provides an efficient route for the synthesis of heteropolynuclear oxalate-bridged species of different nuclearity and dimensionality, and with a variety of magnetic properties [3--5]. 

In this paper magnetic properties of the heterometallic complex [Cu(bpy)$_3$]$_2$-[Cr(C$_2$O$_4$)$_3$]NO$_3\cdot$9H$_2$O (bpy = 2,2'-bipyridine) in the form of single crystals have been investigated. The compound, in which the oxalate anion does not act as a bridging ligand but as a terminal one, was synthesized by the reaction of the [Cr(C$_2$O$_4$)$_3$]$^{3-}$ and [Cu(bpy)$_3$]$^{2+}$ complex ions as building blocks, during the course of a more extensive research work on oxalate-based transition metal species \cite{prvi}. The magnetic measurements performed previously on a powder sample of this heterometallic complex showed a small deviation of the effective magnetic moment from the spin-only value for the three uncoupled spins $(S_{\mathrm{Cu}},S_{\mathrm{Cr}},S_{\mathrm{Cu}})=(1/2,3/2,1/2)$ at low temperatures. This observation could not be well understood in the model of exchange interaction between Cr$^{\mathrm{III}}$ and Cu$^{\mathrm{II}}$ ions. It was suggested that the low temperature magnetic properties of this complex could be explained by considering the zero-field splitting on Cr$^{\mathrm{III}}$. For this purpose, in the present study, the magnetic moment of the single crystals of the complex has been measured. Also the EPR measurements have been performed in order to obtain the spin-Hamiltonian parameters. Magnetization measurements have been analysed with respect to the crystal structure of the compound and the EPR measurements using an exact, instead of the Van Vleck, approach.

\section{Experimental}
Electron paramagnetic resonance (EPR) and magnetization measurements were performed on the single crystals and also on a powdered sample of [Cu(bpy)$_3$]$_2$[Cr(C$_2$O$_4$)$_3$]NO$_3\cdot$9H$_2$O. Transparent-blue single crystals of the complex were obtained by slow evaporation of an aqueous solution containing the [Cr(C$_2$O$_4$)$_3$]$^{3-}$ and [Cu(bpy)$_3$]$^{2+}$ ions in the ratio 1:1 \cite{prvi}. Dimensions of the prepared plate-like single crystals were approximately $2.0 \times 0.4 \times 0.2$ mm$^3$ up to $3.0 \times 0.5 \times 0.5$ mm$^3$. 

For the EPR measurements, the crystal was mounted on a quartz holder in the cavity of an X-band EPR spectrometer (Bruker Elexsys 580 FT/CW) equipped with a standard Oxford Instruments model DTC2 temperature controller. The measurements were performed at the microwave  frequency around 9.6 GHz with the magnetic field modulation amplitude of 3 G at 100 kHz. The EPR spectra were measured at different temperatures, from the room to the liquid helium temperature. It was found that the spectra were approximately temperature independent (except for the paramagnetic behaviour of the spectral line); therefore, only the room temperature spectra are presented in the paper. In the first set of the EPR measurements the crystal was mounted to the quartz holder so that the crystallographic $b$ axis was perpendicular to the magnetic field. The crystal was rotated around the $b$ axis and the EPR spectra were recorded every $5^{\circ}$. Since the complex has a monoclinic crystal symmetry, the $a^{\ast}$ and $c^{\ast}$ axes, mutually perpendicular and perpendicular to the $b$ axis, were chosen arbitrarily. The first rotation was done around the $b$ axis in the $a^{\ast}-c^{\ast}$ plane. For the second set of measurements, the crystal was rotated for $90^{\circ}$ in such a way that it performed rotation in the $a^{\ast}-b$ plane and the third set included the rotation in the $b-c^{\ast}$ plane. Hence, all three rotations were performed around the chosen laboratory axes: $a^{\ast}$, $b$ and $c^{\ast}$ with the magnetic field in the plane of rotation. The rotation was controlled by a goniometer with the accuracy of $0.5^{\circ}$. The largest uncertainty (2--3$^{\circ}$) was related to the optimal deposition of the crystal on the quartz holder.

For the magnetization measurements, a single crystal of the complex was attached to a homogeneous straw using a small dot of vacuum grease. Magnetic moment of the sample was measured using the commercial SQUID magnetometer MPMS-5 (Quantum Design). Measurements were performed in the temperature range 1.8--300 K for the applied magnetic field of $B=1$ T. Also, magnetic moment of the sample in the dependence on the applied magnetic field 0--5.5 T at several temperatures was measured. The measurements were performed for two different orientations of the single crystal with respect to the applied magnetic field: for the magnetic field parallel to the crystallographic \emph{b} axis ($B\parallel b$) and for the magnetic field perpendicular to the crystallographic \emph{b} axis ($B\perp b$). Two crystals were used during the measurements. Mass of the first crystal (for $B\parallel b$) amounted $0.136\pm 0.001$ mg and mass of the second crystal (for $B\perp b$) amounted $0.070\pm 0.005$ mg. Because of the very small masses of the crystals, measurements of the temperature dependence of the magnetic moment were carried out in the magnetic field of 1 T in order to obtain a well measurable signal. The calculated molar magnetizations were corrected for the diamagnetic contribution of the constituent atoms, temperature independent paramagnetism of the Cr$^{\mathrm{III}}$ ion and two Cu$^{\mathrm{II}}$ ions \cite{magnetochemistry} and for a tiny contribution of the background (vacuum grease).

\section{Results and Discussion}

\subsection{Structural considerations}

The details of crystal structure of the complex [Cu(bpy)$_3$]$_2$[Cr(C$_2$O$_4$)$_3$]NO$_3-\cdot$9H$_2$O
(bpy = 2,2'-bipyridine) were described previously \cite{prvi}. The complex crystallizes in the monoclinic space group $P2_1/c$, with the unit cell parameters $a=31.314(3)$ \r{A}, $b=13.5356(11)$ \r{A}, $c=22.202(2)$ \r{A}, $\beta=132.012(13)^{\circ}$ and $Z=4$. The paramagnetic centres (Cr$^{\mathrm{III}}$ in the [Cr(C$_2$O$_4$)$_3$]$^{3-}$ anion and two Cu$^{\mathrm{II}}$ in two [Cu(bpy)$_3$]$^{2+}$ cations) are located in distorted octahedral coordination environments. Distortions are expected to be trigonal (due to the rigidity of the didendate oxalate and bypiridine ligands, respectively) and additionally, in the case of [Cu(bpy)$_3$]$^{2+}$ cations, tetragonal (due to the Jahn-Teller effect). The tetragonal component of the [Cu(bpy)$_3$]$^{2+}$ cation distortion can be easily recognized in the differences of the Cu--N bond lengths as described in \cite{prvi}. For the ligand field environment of the Cr$^{\mathrm{III}}$ ion the tetragonal distortion is negligible as seen from the very small fluctuations of the Cr--O bond lengths ($\sim 0.02$ \r{A}). Instead, rather equal values of the three bite angles ($\sim 82.7^{\circ}$) show a trigonal type of distortion; bite angles are defined as the angles between two coordinated oxygen atoms from the same oxalate ligand and the Cr$^{\mathrm{III}}$ ion. Geometrical parameters for the trigonal distortion (compression ratio, $s/h$ and angle of twisting, $\phi$ \cite{bero}) can be estimated from the positions of six coordinated oxygen atoms: O1, O2, O5, O6, O9 and O10 (Fig. 1(a)). Assuming that the [Cr(C$_2$O$_4$)$_3$]$^{3-}$ anion has an approximate $D_3$ symmetry and that a local three-fold rotation axis passes through the centres of the triangles (O1, O5, O9) and (O2, O6, O10), the values for $s/h$ and $\phi$ are found to be 1.32(3) and 52.2(7)$^{\circ}$, respectively, quite different from the ideal octahedral values of 1.22 and 60$^{\circ}$ \cite{bero}. The mentioned three-fold rotation axis is the only one rotation symmetry of the [Cr(C$_2$O$_4$)$_3$]$^{3-}$ anion which transfers the oxalate groups into each other. 

\begin{figure}[!h]
\centering
\includegraphics[scale=1]{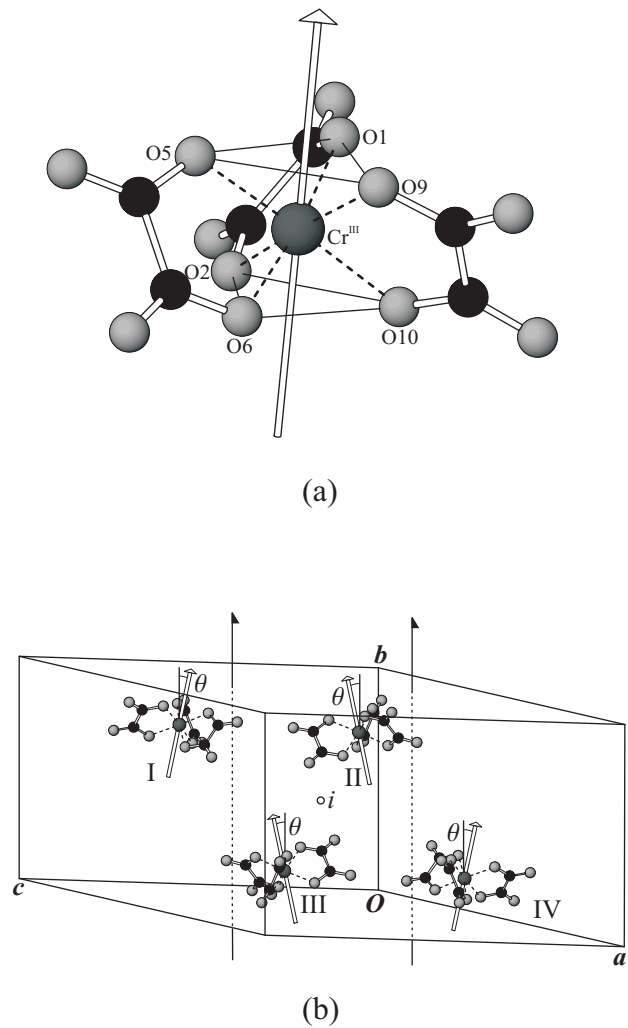}
\caption{(a) The local, approximate three-fold rotation axis of the [Cr(C$_2$O$_4$)$_3$]$^{3-}$ anion (white rod with a triangle on the top) which passes through the Cr$^{\mathrm{III}}$ ion and through the centres of two triangles of coordinated oxygen atoms:  (O1, O5, O9) and (O2, O6, O10).
(b) Four symmetry related [Cr(C$_2$O$_4$)$_3$]$^{3-}$ anions (I, II, III and IV) in the unit cell of the investigated compound showing the inclination angle $\theta$ of their local three-fold rotation axes with respect to the crystallographic $b$ axis. Two two-fold screw symmetry axes and the centre of inversion $i$ are also shown. The [Cu(bpy)$_3$]$^{2+}$ cations and other constituents of the complex are omitted for the reasons of clarity.}
\label{os}
\end{figure}

As a consequence of the symmetry operations in the $P2_1/c$ space group, four [Cr(C$_2$O$_4$)$_3$]$^{3-}$ anions are located in one unit cell (labeled as I, II, III and IV in Fig. 1(b)). Local three-fold rotation axes of [Cr(C$_2$O$_4$)$_3$]$^{3-}$ anions are inclined for $\theta=11.77(1)^{\circ}$ with respect to the crystallographic $b$ axis. The anions I and IV (as well as the anions II and III) have identical orientation in the space because they are related through the centre of inversion $i$ (shown in Fig. 1(b)). The anions related through the two-fold screw symmetries (I and III or II and IV) have mutually different orientation in space as a consequence of the above mentioned angle of $\theta=11.77(1)^{\circ}$ between the local three-fold rotation axes and the crystallographic $b$ axis ($b$ axis is parallel with the two-fold screw symmetry axes). Therefore, the local three-fold rotation axes of the anions I and IV make an angle of $23.54(1)^{\circ}$ with respect to the local three-fold rotation axes of the anions II and III (Fig. 1(b)).

\subsection{EPR study}

\begin{figure}[!h]
\centering
\includegraphics[scale=0.9]{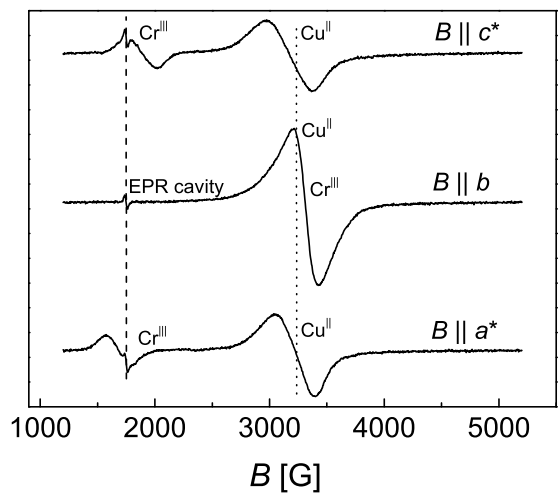}
\caption{The EPR spectra of a single crystal of the investigated complex recorded at room temperature, with the magnetic field $B$ parallel to the laboratory $a^{*}$, $b$ and $c^{*}$ axes.}
\label{epr}
\end{figure}

The EPR spectra of a single crystal of the [Cu(bpy)$_3$]$_2$[Cr(C$_2$O$_4$)$_3$]NO$_3\cdot$9H$_2$O complex when the magnetic field was parallel to the $a^{*}$, $b$ and $c^{*}$ axes are shown in Fig. \ref{epr}. 
In the spectra two types of signals can be separated: the first one from Cu$^{\mathrm{II}}$ appearing always near 3232 G and the second one from Cr$^{\mathrm{III}}$ which shows a large angular dependence: from $\sim 1700-1800$ G $(B \parallel a^{*}$ and $B \parallel c^{*})$ to $\sim 3300$ G $(B\parallel b)$ and even splitting into two lines when the crystal was rotated around the $a^{*}$ and $c^{*}$ axes. The third signal, at 1750 G, comes from the EPR cavity.

\begin{figure}[!h]
\centering
\includegraphics[scale=0.9]{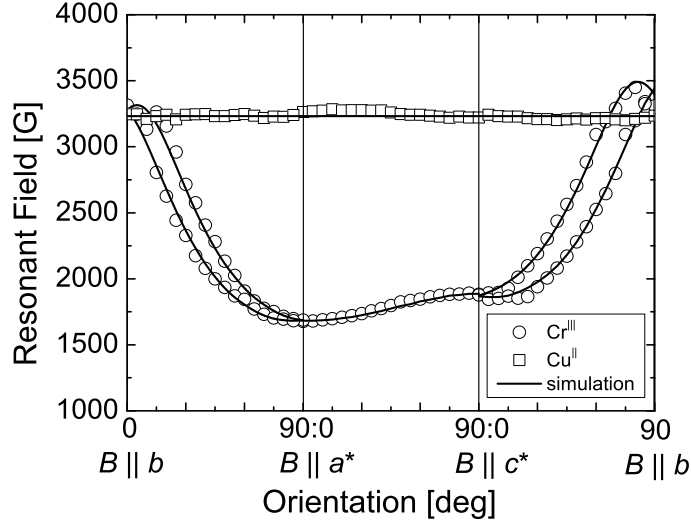}
\caption{Experimental (circles and squares) and simulated (solid lines) angular variation of the resonant field for the single crystal of the investigated complex at room temperature.}
\label{epr2}
\end{figure}

Fig. \ref{epr2} presents the experimental and computer simulated angular dependence of the resonant field for the single crystal of the complex for all three rotations, i.e. for the rotations around the $c^{\ast}$ (from $B \parallel b$ to $B \parallel a^{\ast}$), $b$ (from $B \parallel a^{\ast}$ to $B \parallel c^{\ast}$) and $a^{\ast}$ (from $B \parallel c^{\ast}$ to $B \parallel b$) axes. A computer simulation was carried out by the XSophe-Sophe-XeprView software \cite{hanson}.

The observed spectral lines are very broad ($\sim 350$ G) mainly because of the dipole-dipole interaction so the spectra from Cu$^{\mathrm{II}}$ have been explained approximately by the Zeeman interaction of the electron spin $S=1/2$ with the magnetic field:
\begin{equation}\label{hamcu}
\mathbf{H}=\mu_B \vec{B}\cdot \mathbf{g} \cdot \mathbf{S}
\end{equation}
where $\mu_B$ is the Bohr magneton, $\vec{B}$ is the applied magnetic field, \textbf{g} is  the $g$-tensor and \textbf{S} is the electron spin operator. It has been assumed that the $g$-tensor for Cu$^{\mathrm{II}}$ is isotropic with the $g$-factor value $g=2.11\pm0.02$. In this paper, we have not further analysed the   anisotropy of the $g$-tensor for Cu$^{\mathrm{II}}$ because its resonant field, compared to that from Cr$^{\mathrm{III}}$, depends on orientation only weakly.

For Cr$^{\mathrm{III}}$, the following form of the spin-Hamiltonian has been assumed: 
\begin{equation}\label{ham1}
\mathbf{H}=\mu_B \vec{B}\cdot \mathbf{g} \cdot \mathbf{S}+\mathbf{S}\cdot \mathbf{D} \cdot \mathbf{S}
\end{equation}
where \textbf{D} is the zero-field splitting tensor. The relation (2) could be written equivalently (for more details, see for example [10]):
\begin{equation}\label{ham2}
\mathbf{H}=\mu_B \vec{B}\cdot \mathbf{g} \cdot \mathbf{S} + D \left[ \mathbf{S_z}^2-\frac{S(S+1)}{3} \right ] + E \left ( \mathbf{S_x}^2 - \mathbf{S_y}^2 \right )
\end{equation}
where $D$ and $E$ are the axial and rhombic zero-field splitting parameters, respectively, and the other parameters have their usual meanings. We have assumed an isotropic $g$-tensor for Cr$^{\mathrm{III}}$. A set of parameters reproducing experimental data the best is the following one: $g=1.963\pm0.002$, $D=(0.63\pm0.01)\mbox{ cm}^{-1}$, $\left|E\right|=(0.02\pm0.01)\mbox{ cm}^{-1}$ and the angle between the magnetic $z$ axis and the laboratory $z$ axis (which corresponds to the crystallographic $b$ axis) is $\theta=(11.5\pm0.5)^{\circ}$. This angle is in excellent agreement with the previously mentioned value of $11.77(1)^{\circ}$ for the angle between the calculated three-fold rotation axes of Cr$^{\mathrm{III}}$ and the crystallographic $b$ axis (Sect. 3.1). Two signals from Cr$^{\mathrm{III}}$ observed in the $b-a^{*}$ and $c^{*}-b$ planes (Fig. 3) occur because two orientations of the three-fold rotation axes exist (Fig. 1(b)). The EPR transition for Cr$^{\mathrm{III}}$ which can be seen in Fig. \ref{epr} corresponds to the transition $M_S=-1/2 \leftrightarrow M_S=1/2$. Another allowed transition $M_S=-1/2 \leftrightarrow M_S=-3/2$ should be observed at a higher magnetic field which was out of range of the experimental set-up (behind 1 T) and the transition $M_S=1/2 \leftrightarrow M_S=3/2$ could not be observed by the X-band EPR spectrometer, because the zero field splitting $D=0.63\mbox{ cm}^{-1}\approx 20\mbox{ GHz}$ is much higher than our microwave frequency of 9.6 GHz \cite{carrington}. These EPR measurements could give only an absolute value of the parameter $D$, but the positive sign was chosen in accordance to the magnetization measurements. The sign of $E$ has no physical meaning, except in terms of convention of the chosen $a^{*}$ and $c^{*}$ axes \cite{weil}.

\begin{figure}[!h]
\centering
\includegraphics[scale=0.85]{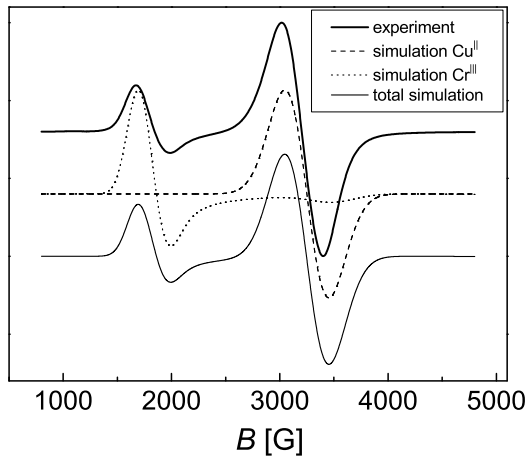}
\caption{The powder EPR spectrum of the investigated complex recorded at room temperature (thick solid line). The simulations for Cu$^{\mathrm{II}}$ (dashed line) and for Cr$^{\mathrm{III}}$ (dotted line) are obtained using the single crystal parameters. The total simulation line (thin solid line) is obtained as a sum of the simulation lines for Cu$^{\mathrm{II}}$ and for Cr$^{\mathrm{III}}$ in a ratio $2:1$.}
\label{powder}
\end{figure}

In order to additionally check the spin-Hamiltonian parameters obtained from the single crystal data, the EPR spectrum of a powdered sample of the complex is recorded at room temperature, Fig. \ref{powder}.

The powder simulation is performed using the previously mentioned parameters for Cu$^{\mathrm{II}}$ and for Cr$^{\mathrm{III}}$ obtained from the single crystal simulation. The total simulation line of the powder spectra is obtained as a sum of the simulation lines for Cu$^{\mathrm{II}}$ and for Cr$^{\mathrm{III}}$ in a ratio $2:1$ according to the fact that there is two times more Cu$^{\mathrm{II}}$ than Cr$^{\mathrm{III}}$ ions in the unit cell. It could be seen from Fig. \ref{powder} that the simulation is in good agreement  with the powder spectrum.

\subsection{Magnetization study}

\begin{figure}[!h]
\centering
\includegraphics[scale=0.85]{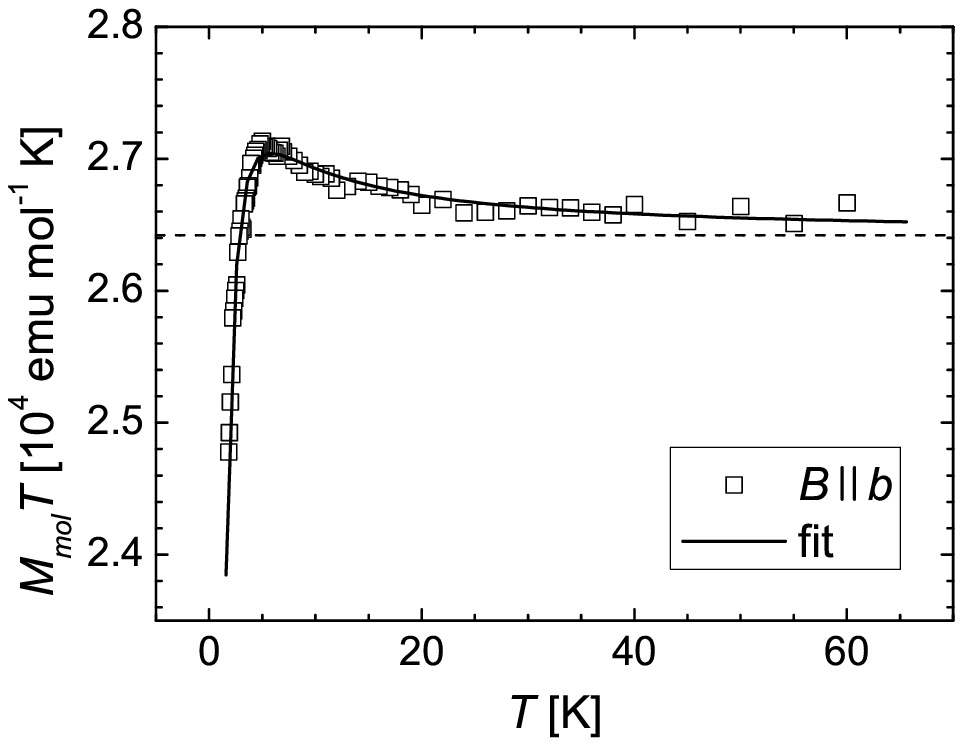}
\caption{The $M_{mol}T$ versus $T$ plot for the parallel orientation of the crystallographic \emph{b} axis with respect to the applied magnetic field $B=1$ T for the investigated complex. The horizontal, dashed line corresponds to the theoretical value expected for the three uncoupled spins.}
\label{slika1}
\end{figure}

\begin{figure}[!h]
\centering
\includegraphics[scale=0.85]{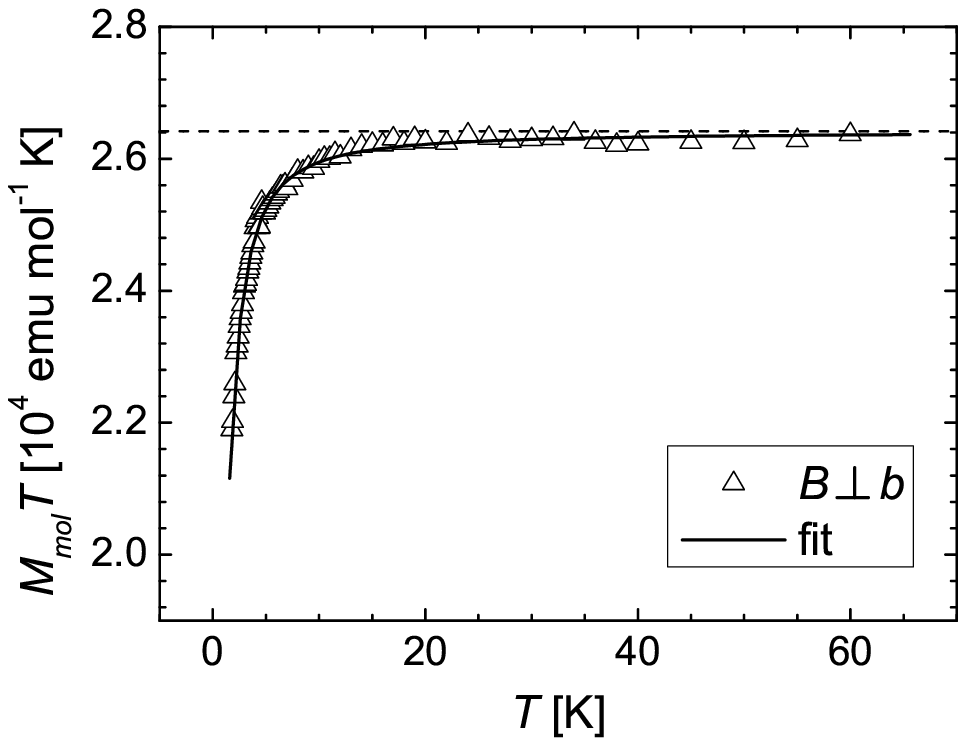}
\caption{The $M_{mol}T$ versus $T$ plot for the perpendicular orientation of the crystallographic \emph{b} axis with respect to the applied magnetic field $B=1$ T for the investigated complex. The horizontal, dashed line corresponds to the theoretical value expected for the three uncoupled spins.}
\label{slika2}
\end{figure}

The results of magnetization measurements of the [Cu(bpy)$_3$]$_2$[Cr(C$_2$O$_4$)$_3$]-NO$_3\cdot$9H$_2$O complex for the parallel and perpendicular orientations of the crystallographic \emph{b} axis with respect to the applied magnetic field $B=1$ T, i.e. for $B\parallel b$ and $B\perp b$, are shown in Figs. \ref{slika1} and \ref{slika2}, respectively, as the dependence of the product of molar magnetization and temperature, $M_{mol}T$ on temperature $T$.

At high temperatures the value of the product of molar magnetization and temperature, $M_{mol}T$ should correspond to the spin-only value.
For the three uncoupled spins $(S_{\mathrm{Cu}},S_{\mathrm{Cr}},S_{\mathrm{Cu}})=(1/2,3/2,1/2)$ this value is given by the expression:
\begin{equation}
M_{mol}T=\frac{N_A \mu_B^2}{3k}[g_{\mathrm{Cu}}^2S_{\mathrm{Cu}}(S_{\mathrm{Cu}}+1)+g_{\mathrm{Cr}}^2S_{\mathrm{Cr}}
(S_{\mathrm{Cr}}+1)+g_{\mathrm{Cu}}^2S_{\mathrm{Cu}}(S_{\mathrm{Cu}}+1)]
\end{equation}
where $N_A$ is Avogadro's number and $k$ is the Boltzmann constant. By using the $g$-factors $g_{\mathrm{Cr}}=1.963$ and $g_{\mathrm{Cu}}=2.11$ (obtained from the EPR measurements) the $M_{mol}T$ product equals $2.642 \cdot 10^4$ emu mol$^{-1}$ K (shown as the dashed line in Figs. \ref{slika1} and \ref{slika2}). The value of the $M_{mol}T$ product for $B\parallel b$ (Fig. \ref{slika1}) is constant at high temperatures and corresponds to the predicted spin-only value. The $M_{mol}T$ product increases on cooling, attains a maximum value at 5 K and thereupon decreases. For $B\perp b$ (Fig. \ref{slika2}), the $M_{mol}T$ product is also constant at high temperatures, having the value as expected for the three independent spins. On cooling, this value remains constant down to 15 K and decreases on further cooling. The relative deviation of the $M_{mol}T$ value at low temperatures from the spin-only value is rather small, i.e. approximately 9\% for $B\parallel b$ and 17\% for $B\perp b$.

In the previous investigation of magnetic properties of this heterometallic complex \cite{prvi}, it was observed that the long-range intermolecular interaction was negligible. Also, exchange interaction between Cu$^{\mathrm{II}}$ and Cr$^{\mathrm{III}}$ ions and between two Cu$^{\mathrm{II}}$ ions can be discarded because in the investigated heterometallic complex Cu$^{\mathrm{II}}$ and Cr$^{\mathrm{III}}$ ions are not bridged by diamagnetic ligands which could mediate the exchange interaction. Besides, by using the model of exchange interaction between paramagnetic centres it was not possible to explain the decrease of  the $M_{mol}T$ product at low temperatures and the observed magnetic anisotropy \cite{prvi}. Therefore, in further analysis of magnetic properties of the complex we have excluded exchange interaction between the magnetic ions. Considering the EPR measurements, we have tried to associate the observed magnetic anisotropy at low temperatures to the zero-field splitting of Cr$^{\mathrm{III}}$ ion levels in a distorted octahedral surrounding ($g$-factors for Cr$^{\mathrm{III}}$ and Cu$^{\mathrm{II}}$ ions assumed to be isotropic).

With this aim in view, in this paper we have derived the exact expression for the molar magnetization of the complex. The Van Vleck approach has not been used because the condition $\mu_B B/kT \ll 1$ is not fulfilled at low temperatures and for the applied magnetic field 1 T. Moreover, the deviation of the $M_{mol}T$ product from the spin-only value appears at relatively low temperatures and is relatively small. Also, with regard to the EPR results and other studies of similar heterometallic complexes in which Cr$^{\mathrm{III}}$ ions also adopt distorted octahedral coordination \cite{zfs2,zfs3}, it is reasonable to expect that the value of the axial zero-field splitting parameter is of the order 1 cm$^{-1}$. Thus, neither the condition $D \gg g_{\mathrm{Cr}} \mu_B B_{\perp}$, usually used in the Van Vleck formula, is valid for the applied magnetic field of 1 T. In literature where zero-field splitting is studied, commonly the Van Vleck formula is used with an additional energy linearization \cite{kahn}. In the present investigation the idea has been to find out to what extent the approximations used in the Van Vleck approach have the influence on the interpretation of magnetic measurements and on the quantitative determination of parameter $D$, too.

Generally, molar magnetization of a quantum system with an energy spectrum $E_n$, $n=1,2,3, \ldots$ in the presence of a magnetic field $B$ in thermal equilibrium is given by the Boltzman distribution law:
\begin{equation}\label{magnetizacija}
M_{mol}=\frac{N_A \sum_n \limits - \frac{\partial E_n}{\partial B} e^{-E_n/kT}}{\sum_n \limits e^{-E_n/kT}}
\end{equation}
The equation (\ref{magnetizacija}) is the fundamental expression for magnetization and does not lean on any approximation. Energy spectrum $E_n$, $n=1,2,3, \ldots$ corresponds to the eigenvalues of the Hamiltonian of the given system. 

\begin{figure}[!h]
\centering
\includegraphics[scale=0.9]{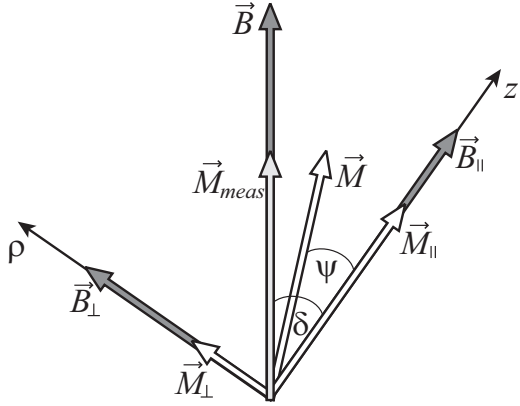}
\caption{Schematic view of the magnetic field and magnetization vectors, $\vec{B}$ and $\vec{M}$, respectively, when $\vec{B}$ is applied in an arbitrary direction to the anisotropy axis $z$. $B_{\parallel}$ and $B_{\perp}$ are components of $\vec{B}$ ($M_{\parallel}$ and $M_{\perp}$ are components of $\vec{M}$) parallel and perpendicular to the anisotropy axis, respectively; $\vec{M}$ is the resultant magnetization and $M_{meas}$ is a projection of the magnetization to the direction of the applied field measured in the experiment.}
\label{zfs}
\end{figure}

The Hamiltonian of the Cr$^{\mathrm{III}}$ ion in a distorted octahedral surrounding in the presence of the magnetic field is given by the relation (\ref{ham2}), which can be written in an equivalent form:
\begin{equation}\label{ham}
\mathbf{H}=g_{\mathrm{Cr}} \mu_B \mathbf{S_z} \cdot B_{\parallel} + g_{\mathrm{Cr}} \mu_B \mathbf{S_x} \cdot B_{\perp} + D \left[ \mathbf{S_z}^2-\frac{S(S+1)}{3} \right ] + E \left ( \mathbf{S_x}^2 - \mathbf{S_y}^2 \right )
\end{equation}
where $B_{\parallel}$ and $B_{\perp}$ are components of the applied magnetic field parallel and perpendicular to the anisotropy axis of the local surrounding of the Cr$^{\mathrm{III}}$ ion, respectively, the $g$-tensor is assumed to be isotropic with the $g_{\mathrm{Cr}}$ principal value and the other symbols have their usual meanings. The energy levels $E_{n,\mathrm{Cr}}$ are determined by diagonalization of the above Hamiltonian. The components of molar magnetization parallel $M_{\parallel}$ and perpendicular $M_{\perp}$ to the anisotropy axis are given by the following equation:
\begin{equation}\label{mzx}
M_{\parallel,\perp;\mathrm{Cr}}=N_A \frac{\sum_n \limits - \frac{\partial E_{n,\mathrm{Cr}}}{\partial B_{\parallel,\perp}} e^{-E_{n,\mathrm{Cr}}/kT}}{\sum_n \limits e^{-E_{n,\mathrm{Cr}}/kT}}
\end{equation}
Generally, when the magnetic field $B$ is applied in an arbitrary direction (i.e. neither parallel nor perpendicular to the anisotropy axis), the resultant magnetization $M$ is not parallel to the applied field $B$. The magnetic field vector $\vec{B}$ and the magnetization vector $\vec{M}$  with its parallel and perpendicular components for such a general case are shown in Fig. \ref{zfs}.

The total magnetization is a vector sum of the $M_{\parallel}$ and $M_{\perp}$ components.
As presented in Fig. \ref{zfs}, $\delta$ is the angle between the applied magnetic field $B$ and the anisotropy axis $z$ and $\psi$ is the angle between the total magnetization $M$ and the anisotropy axis $z$.
Projection of the total magnetization to the direction of the applied magnetic field is measured in the experiment ($M_{meas}$). Therefore, contribution of the Cr$^{\mathrm{III}}$ ion to the molar magnetization of the complex is given by the following expression:
\begin{equation}
M_{\mathrm{Cr}}(B,T,\delta)=\sqrt{M_{\parallel,\mathrm{Cr}}^2+M_{\perp,\mathrm{Cr}}^2} \cdot \cos{\left[ \delta - \arctan \left({\frac{M_{\perp,\mathrm{Cr}}}{M_{\parallel,\mathrm{Cr}}}}\right ) \right ]}
\end{equation}

In addition to the contribution of the Cr$^{\mathrm{III}}$ ion to the molar magnetization of the compound, there is also a paramagnetic contribution of the two Cu$^{\mathrm{II}}$ ions. The Hamiltonian of the Cu$^{\mathrm{II}}$ ion in an external magnetic field is given by (\ref{hamcu}) and its eigenvalues $E_{n,\mathrm{Cu}}$ are used to evaluate molar magnetization of the Cu$^{\mathrm{II}}$ ion according to the equation (\ref{magnetizacija}).

In the experiment the magnetic field $\vec{B}$ was parallel or perpendicular to the crystallographic $b$ axis. In these two specific cases both anisotropy axes (i.e. local three-fold rotation axes) form the same angle $\delta$ with the magnetic field, as it can be observed in Fig. \ref{os}(b). The final expression for the measured magnetization when the magnetic field is parallel or perpendicular to the crystallographic $b$ axis is:
\begin{equation}\label{mmj}
M_{meas}(B,T,\delta)=M_{\mathrm{Cr}}(B,T,\delta) +2M_{\mathrm{Cu}}(B,T).
\end{equation}

The nonlinear fitting of the measured $M_{mol}T$ data versus temperature dependence in the temperature range 1.8--60 K for both crystals was performed by the equation (\ref{mmj}) multiplied by temperature $T$. Fitting parameters were the axial zero-field splitting parameter $D$, the angle between the anisotropy axes and the magnetic field $\delta$ and a multiplication constant taking into account uncertainty of the weighed mass of the sample.
The values of the $g$-factors were fixed to $g_{\mathrm{Cr}}=1.963$ and $g_{\mathrm{Cu}}=2.11$ and the rhombic zero-field splitting parameter to $E=0$. The nonlinear fitting gave the following values: $D=1.03$ cm$^{-1}$, $\delta=69.9^{\circ}$ for the parallel orientation and $D=0.97$ cm$^{-1}$, $\delta=48.2^{\circ}$ for the perpendicular orientation. The discrepancy factor, defined as $R=\sum \left [ (M_{mol}T)_{obs}-(M_{mol}T)_{calc} \right ] ^2/ \sum \left [ (M_{mol}T)_{obs} \right ] ^2$ equaled $5.7 \cdot 10^{-6}$ and $9.0 \cdot 10^{-6}$ for the parallel and perpendicular orientations,
respectively. The obtained parameters $D$ are consistent for both orientations. According to the definiton and geometrical reasons, the angle $\delta$ in the $B\perp b$ case should be in the interval $[90^{\circ}-\delta(B\parallel b), 90^{\circ}]$. The obtained values of the angle $\delta$ in both cases fulfill this requirement.
The calculated curves for $B\parallel b$ and $B\perp b$ are presented in Figs. \ref{slika1} and \ref{slika2}, respectively, showing good overlapping with the measured data.

Several additional measurements of the temperature dependence of the magnetic moment of the complex for $B\parallel b$ and $B\perp b$ were performed on different single crystals. The nonlinear fitting gave mutually consistent values of the parameters $D$ and $\delta$.

For the $B \parallel b$ case the angle $\delta$ should correspond to the previously mentioned angle between crystallographic $b$ axis and the anisotropy axis $\theta$, which is determined from the crystallographic and EPR data (Section 3.1 and 3.2). The calculated curve with the parameters obtained by the nonlinear fitting of the $M_{mol}T$ data in the $B \parallel b$ case with the value of the angle $\delta$ fixed to the value of $\theta=11.77^{\circ}$ was not in accordance with the measured data. So, the angle $\delta$ is used as one of the fitting parameters. In the $B \perp b$ case the value of the angle $\delta$ could not be predicted from the structural considerations. Therefore, in the nonlinear fitting in the $B \perp b$ case $\delta$ is also taken as fitting parameter.

\begin{figure}[!h]
\centering
\includegraphics[scale=0.85]{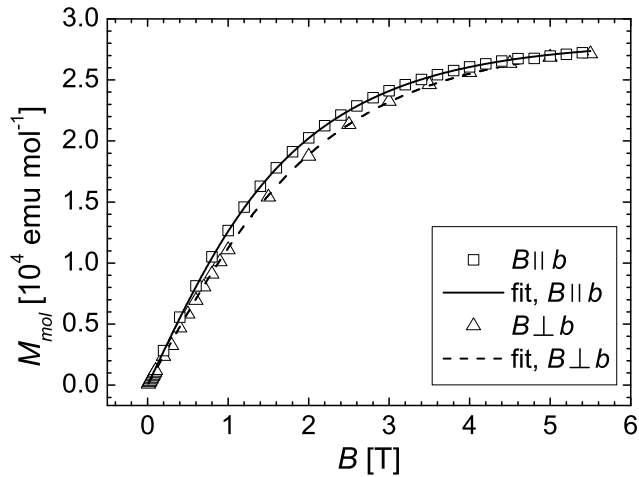}
\caption{The $M_{mol}$ versus $B$ plot for the parallel and perpendicular orientations of the single crystals of the investigated complex with respect to the applied magnetic field at 2 K.}
\label{MH}
\end{figure}

The nonlinear fitting of the molar magnetization $M_{mol}$ versus magnetic field $B$ dependence in the range 0--5.5 T was performed for the measurements at 2 K. It can be shown that the shape of the $M_{mol}(B)$ curve is rather insensitive to the values of $D$ and $\delta$. So, all parameters as before were used except the multiplicative constant. The results of the fit have shown that the measured data can be well described by the relation (\ref{mmj}). The experimental data and the calculated curves of $M_{mol}(B)$ for $B\parallel b$ and $B \perp b$ are given in Fig. \ref{MH}. They clearly show that a linear approach in magnetic susceptibility calculation would be meaningless for the used ranges of magnetic field and temperature.

In the Van Vleck approach (described in \cite{kahn}) it is assumed that the energy levels $E_n$ can be expanded according to the increasing powers of $B$ ($E_n=E_n^{(0)}+E_n^{(1)}B+E_n^{(2)}B^2+\cdots$) and that the condition $B/kT \ll 1$ is fulfilled. Introducing these ap
The nonlinear fitting of the formula obtained by the Van Vleck approach to the experimental data gave about two times higher values of the axial zero-field splitting parameter $D$, which is also in bigger discrepancy with the results of the EPR measurements. Fit to the experimental data with the parameter $E$ as fifth fit parameter did not give physical and reliable results because of overparametrization and other numerical problems.

The values of the parameter $D$ obtained from two independent measurements, that is for two different orientations of the single crystal in the magnetic field are consistent to each other, and temperature dependence of the magnetization of the complex can be well described by this model. Uncertainty of the obtained parameters can be estimated in the following manner. First, it should be emphasized that the effect of zero-field splitting is observable only at low temperatures and that the deviation of the measured magnetization from the spin-only value is rather small. Therefore, the uncertainty of the mass of the sample as well as the inability of the exact determination of the contribution of the background to the measured magnetization contribute nonnegligibly to the uncertainty of the obtained parameters. The influence of the background has been estimated in such a manner that the molar magnetization at high temperatures corresponds to the spin-only value. From the nonlinear fitting of the measured data corrected for the slightly different values of the background magnetization it has been observed that the value of the parameter $D$ is changing within $\pm 0.2$ cm$^{-1}$ and the value of $\delta$ within $\pm 1^{\circ}$, whereas the discrepancy factor $R$ does not change significantly.

The EPR measurements confirm the results of the measurements performed on SQUID magnetometer taking into account associated uncertainties of the calculated parameters and the specificities of the used experimental methods. EPR measurements are suitable for accurate determination of the spin-Hamiltonian parameters of a system. Magnetization measurements can provide only an effective value of the zero-field splitting parameter. Although the zero-field splitting of the energy levels of the Cr$^{\mathrm{III}}$ ion determines mainly the low temperature behaviour of the magnetization of the complex, it is possible that there are some other, much smaller effects which also contribute to the magnetization and may influence the calculated values of the parameter $D$ and angle $\delta$. Hence, the anisotropy of the $g$-factors of the Cr$^{\mathrm{III}}$ and Cu$^{\mathrm{II}}$ ions, the exchange interaction between Cr$^{\mathrm{III}}$ and Cu$^{\mathrm{II}}$ ions and between two Cu$^{\mathrm{II}}$ ions are likely present to a certain extent and possibly have an effect on the calculated value of the parameters. Besides, the magnetization has been measured in the relatively high field (1 T) which may influence the magnitude of anisotropy and the direction of the anisotropy axis. Magnetostrictive effects in such a structure are not known, but it could be possible that the applied field changes the direction and the amount of anisotropy or even produces some additional contribution \cite{LaSrCoO}. This could be the reason of differing between the magnetization and EPR determinations of the zero-field splitting parameter $D$ and of the angle between crystallographic $b$ axis and the anisotropy axis $\theta$. 

\section{Conclusion}

Magnetic properties of the heterometallic complex [Cu(bpy)$_3$]$_2$[Cr(C$_2$O$_4$)$_3$]-NO$_3\cdot$9H$_2$O (bpy = 2,2'-bipyridine) in the form of
the single crystal have been investigated. At low temperatures magnetic
anisotropy has been observed. Taking into consideration the results \cite{prvi} of the previous magnetic measurements and known crystal structure of the complex it has been concluded that the exchange interaction between Cr$^{\mathrm{III}}$ and Cu$^{\mathrm{II}}$ ions and between two Cu$^{\mathrm{II}}$ ions, as well as the long-range intermolecular interaction are negligible. The zero-field splitting of the energy levels of the Cr$^{\mathrm{III}}$ ion in the trigonally distorted octahedral surrounding leads to the magnetic anisotropy at low temperatures. This effect also completely determines the low temperature behaviour of the magnetization of the complex because other possible interactions are negligible, unlike to similar heterometallic complexes where other interactions, such as exchange interaction between paramagnetic centres or long-range interaction dominate. Therefore, this complex is very suitable for qualitative and quantitative studies of zero-field splitting. The theoretical expression for molar magnetization of the complex has been derived without use of any approximation. The nonlinear fitting to the experimental data has shown that the magnetic properties of the compound can be well described by this model.

The calculated value of the axial zero-field splitting parameter of the Cr$^{\mathrm{III}}$ ion is in agreement with those previously reported for similar polynuclear complexes \cite{zfs2,zfs3}. Also, the sign of the parameter $D>0$ has been uniquely determined. Besides the magnetic moment measurement of the single crystals of the complex, the behaviour of the EPR spectra by rotating a single crystal about the three independent axes has been studied. From the spectral line shifting the spin-Hamiltonian parameters have been obtained, which are in agreement with the data from the literature [15--17]. It has been also shown that the Van Vleck approach gives bigger disagreement between two methods. The EPR measurements, which are more relevant for microscopic parameters determination, thus have confirmed the parameters obtained from the magnetic bulk measurements. It has been shown that magnetic measurements are also appropriate for the anisotropy parameter determination if appropriately analysed, i.e. if using exact calculation of the magnetization of a system.

\bigskip
\textbf{Acknowledgments}
\bigskip

We are grateful to Ivan Kup\v{c}i\'{c} for valuable discussions. This research was supported by the Ministry of Science, Education and Sports of the Republic of Croatia (projects 119-119458-1017, 098-0982915-2939 and 098-0982904-2946).

\end{document}